\documentclass[aps,prb,twocolumn,superscriptaddress,showpacs]{revtex4-1}

\usepackage{amssymb,amsmath}
\usepackage[sort&compress]{natbib}
\usepackage{multirow}
\usepackage{color}
\usepackage{graphicx}
\usepackage{dcolumn}
\usepackage{bm}

\begin{document}

\title{Magnetic and Structural Phase Transitions in the Spinel Compound Fe$_{1+x}$Cr$_{2-x}$O$_{4}$}

\author{J. Ma}
\affiliation{Quantum Condensed Matter Division, Oak Ridge National Laboratory, Oak Ridge, TN 37831, USA}

\author{V. O. Garlea}
\affiliation{Quantum Condensed Matter Division, Oak Ridge National Laboratory, Oak Ridge, TN 37831, USA}

\author{A. Rondinone}
\affiliation{Center for Nanophase Materials Sciences, Oak Ridge National Laboratory, Oak Ridge, TN 37831, USA}

\author{A. A. Aczel}
\affiliation{Quantum Condensed Matter Division, Oak Ridge National Laboratory, Oak Ridge, TN 37831, USA}

\author{S. Calder}
\affiliation{Quantum Condensed Matter Division, Oak Ridge National Laboratory, Oak Ridge, TN 37831, USA}

\author{C. dela Cruz}
\affiliation{Quantum Condensed Matter Division, Oak Ridge National Laboratory, Oak Ridge, TN 37831, USA}

\author{R. Sinclair}
\affiliation{Department of Physics and Astronomy, University of Tennessee, Knoxville, TN 37996, USA}

\author{W. Tian}
\affiliation{Quantum Condensed Matter Division, Oak Ridge National Laboratory, Oak Ridge, TN 37831, USA}

\author{Songxue Chi}
\affiliation{Quantum Condensed Matter Division, Oak Ridge National Laboratory, Oak Ridge, TN 37831, USA}

\author{A.~Kiswandhi}
\affiliation{National High Magnetic Field Laboratory, Florida State University, Tallahassee, FL 32306, USA}
\affiliation{Department of Physics, Florida State University, Tallahassee, FL 32306-3016, USA}

\author{J.~S.~Brooks}
\affiliation{National High Magnetic Field Laboratory, Florida State University, Tallahassee, FL 32306, USA}
\affiliation{Department of Physics, Florida State University, Tallahassee, FL 32306-3016, USA}

\author{H. D. Zhou}
\affiliation{Department of Physics and Astronomy, University of Tennessee, Knoxville, TN 37996, USA}
\affiliation{National High Magnetic Field Laboratory, Florida State University, Tallahassee, FL 32306, USA}

\author{M. Matsuda}
\affiliation{Quantum Condensed Matter Division, Oak Ridge National Laboratory, Oak Ridge, TN 37831, USA}
\date{\today}


\begin{abstract}

Neutron and X-ray diffraction, magnetic susceptibility, and specific heat measurements have been used to investigate the magnetic and structural phase transitions of the spinel system Fe$_{1+x}$Cr$_{2-x}$O$_{4}$ ($0.0$ $\le$ $x$ $\le$ $1.0$). The temperature versus Fe concentration ($x$) phase diagram features two magnetically ordered states and four structural states below 420 K. The complexity of the phase diagram is closely related to the change in the spin and orbital degrees of freedom induced by substitution of Fe ions for Cr ions. The systematic change in the crystal structure is explained by the combined effects of Jahn-Teller distortion, spin-lattice interaction, Fe$^{2+}$-Fe$^{3+}$ hopping, and disorder among Fe$^{2+}$, Fe$^{3+}$, and Cr$^{3+}$ ions.

\end{abstract}


\maketitle
\section{Introduction}

Due to the strong interactions among the spin, orbital, and lattice degrees of freedom, the transition metal oxides with spinel structure, $AB_2$O$_4$, present complicated magnetic and structural phase transitions and have attracted extensive attention in past years.\cite{SHLee1, Bramwell, Ramirez} In the system, the octahedrally coordinated $B$-site cations form a geometrically frustrated network of corner shared tetrahedra, while the $A$-site cations form a diamond lattice and are located at the center of oxygen-tetrahedra. The 3$d$ orbitals of the $B$-site cation split into the triply degenerate low-energy $t_{2g}$ states and doubly degenerate high-energy $e_{g}$ states, while the $A$-site cation has two low energy $e_{g}$ states with three high energy $t_{2g}$ states.\cite{Ramirez, Kino} Since the properties of both $A$ and $B$ cations are driven by the electron occupancies on 3$d$ orbitals, which determine the magnetic and orbital degrees of freedom, it is challenging to obtain the original driving forces for those magnetic and structural phase transitions in the spinel oxides.

In order to get insightful information on the phase transition mechanism, one can occupy the $B^{3+}$-site with a spin only cation, (such as the chromite spinels $A$Cr$_2$O$_4$). The electronic configuration for Cr$^{3+}$ cation is 3$d^3$($S$(Cr$^{3+}$)=3/2), which leads to half filled $t_{2g}$ and empty $e_{g}$ orbitals. In addition, it has been found that different $A$-site cations could yield different ordered states: i) If $A^{2+}$ ions are magnetically neutral, such as ZnCr$_2$O$_4$,\cite{SHLee, SHLee1, Kemei, Sushkov, SHLee2} MgCr$_2$O$_4$,\cite{Kemei, Mamiya, SHLee2} CdCr$_2$O$_4$,\cite{SHLee2, Chung} and HgCr$_2$O$_4$,\cite{Ueda, Matsuda, SHLee2} a transition from the paramagnetic cubic phase to the N\'{e}el-ordered tetragonal or orthorhombic one at low temperatures is obtained. ii) If $A^{2+}$ ions are magnetic with spin only, such as MnCr$_2$O$_4$ ($S$(Mn$^{2+}$)=5/2),\cite{Hastings, Tomiyasu, Ederer, Lyons, Kocsis} and CoCr$_2$O$_4$($S$(Co$^{2+}$)=3/2),\cite{Hastings, Menyuk, Plumier, Tomiyasu, Ederer, Lyons, Bordacs, Kocsis, Singh} the lattice remains cubic, with a paramagnetic-to-ferrimagnetic transition at high temperature, followed by a transition to spiral ordering at lower temperature due to weak magnetic geometrical frustration. iii) If $A^{2+}$ ions are magnetic with the orbital degree of freedom, such as FeCr$_2$O$_4$, \cite{Tomiyasuf, Tanaka, Shirane, Bordacs, Kocsis, Singh, Ohgushi} NiCr$_2$O$_4$,\cite{Tomiyasuf, Tomiyasu1, Suchomel, Bordacs, Kocsis, Ohgushi} and CuCr$_2$O$_4$,\cite{Suchomel, Prince, Bordacs, Kocsis, Ohgushi, Jo} a cubic-tetragonal phase transition is observed at a higher temperature, followed by magnetic order, which indicates that the magnetic ordering is stabilized by reducing the lattice symmetry through a spin-lattice coupling. The long-range ordered collinear ferrimagnetic state can eventually evolve into different noncollinear ferrimagnetic states at a lower temperature, such as conical ordering in FeCr$_2$O$_4$ and NiCr$_2$O$_4$, and Yafet-Kittel-type magnetic ordering in CuCr$_2$O$_4$. Moreover, the multiferroic ordering and the dielectric response induced by the magnetic field also have been found in several chromium spinel oxides.\cite{Bordacs, Kocsis, Singh, Ohgushi, Yamasaki, Mufti} Substantial experimental and theoretical works have been performed to study the intriguing properties of $A$Cr$_2$O$_4$. Previously, a largely separate line of research has been devoted to the spin-lattice interaction which is related to the spin frustration and the cooperative Jahn-Teller distortion especially for the compounds involving orbitally active $A$-site cations.\cite{Bordacs, Kocsis, Singh, Suchomel, Ohgushi, Mufti, Kataoka, Katsura, Ohtani, Sagayama} However, very few materials have been studied from the view point of the coupling between frustration and Jahn-Teller effects by changing the orbital configuration of $B$-site cation.

\begin{figure*}
 \centering
  \includegraphics[width=1.0\textwidth]{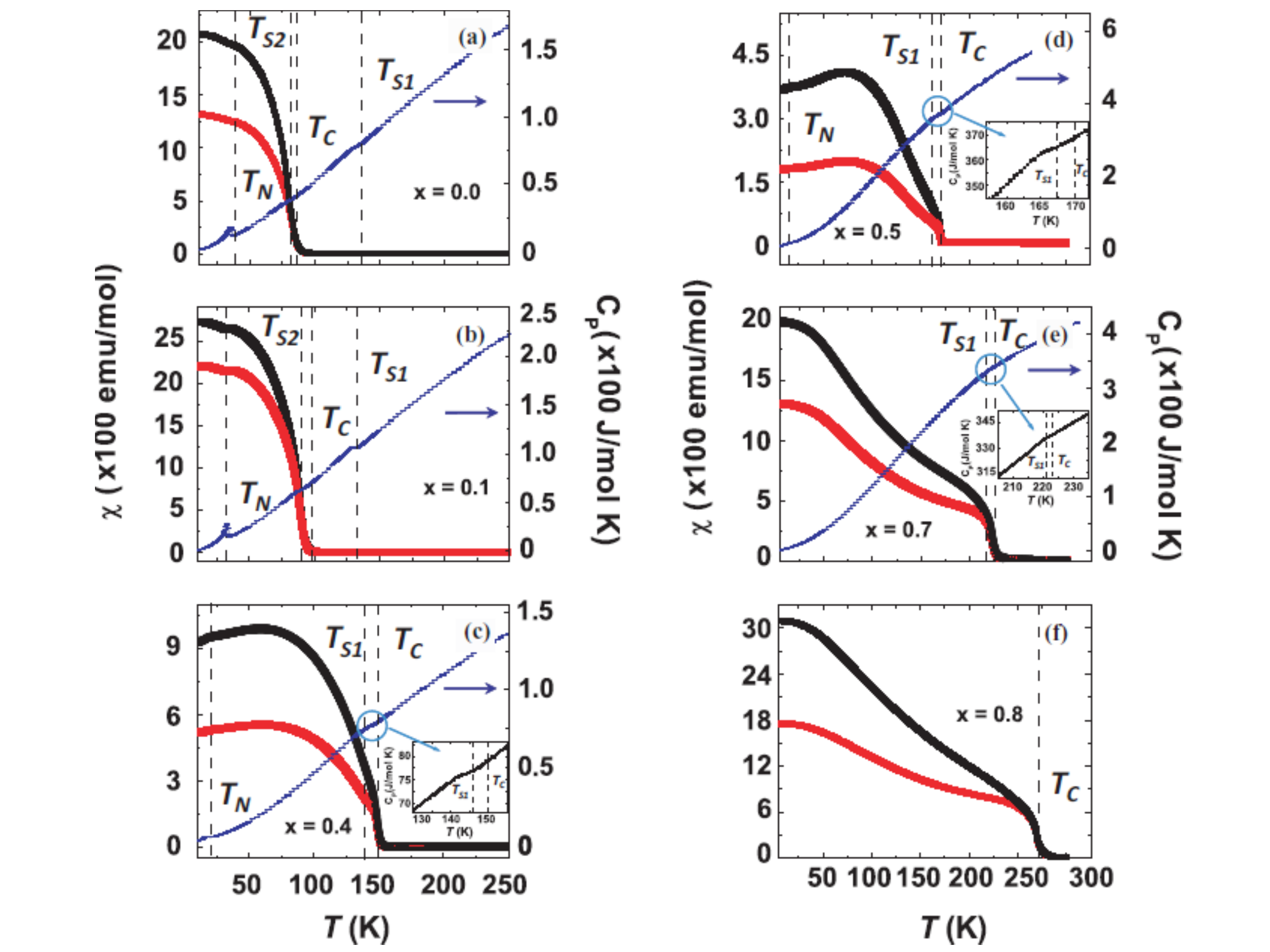}\\
\caption{(Color online) The temperature dependence of magnetic susceptibilities and specific heat for Fe$_{1+x}$Cr$_{2-x}$O$_4$ (0.0 $\le$ $x$ $\le$ 0.8). Black and red lines are the results of the field-cooled(FC) and zero-field-cooled(ZFC) measurements, respectively. The blue line presents the specific heat data. For $x \le$ 0.3, $T_{S_2}$ is close to $T_{C}$; For 0.3$\le x \le$ 0.8, $T_{S_1}$ is close to $T_{C}$. Insets present the enlarged view of the specific heat data around $T_{S_1}$ and $T_C$.}
  \label{SQUID}
\end{figure*}

In this regard, Fe$_{1+x}$Cr$_{2-x}$O$_4$ is a remarkable compound. The Fe$^{3+}$ ions (3$d^5$) are not orbitally active and have a large spin, $S$ = 5/2, while Fe$^{2+}$ ions have an orbital degree of freedom with 3$d^6$ and $S$ = 2. Since the Cr$^{2+}$ energy level lies well above the Fe$^{2+/3+}$ energy levels, the valence of Cr$^{3+}$ is stable. Although electrons have been reported to be hopping between $A$- and $B$-sites for $x$ $\ge$ $\sim$ 0.8, \cite{Yearian, Robbins} the Cr$^{3+}$ ions always stay at the $B$-site and the only electron transfer should be between Fe$^{2+}$ and Fe$^{3+}$ ions. This arrangement leads to the 3$d$ electronic ground state of Fe$^{2+}$ ion changing from $e_g$ on the $A$-site to $t_{2g}$ on the $B$-site and the type of the Jahn-Teller distortion effect being alternated in the system.\cite{Verwey, Yearian, Goodenough, Kyono} Therefore, the primary effect of Fe-doping in this system is to change the average moment on the $B$-sites and alter the competition of antiferromagnetic interactions between $A$-$B$ and $B$-$B$. For this reason, studying the magnetic and structural lattice of Fe$_{1+x}$Cr$_{2-x}$O$_4$ can help to uncover the origin of the orbital ordering effect on the structural transition.

The structure was first discussed by Verwey et al. in 1947,\cite{Verwey} then several techniques had been applied to study the physical properties. In 1964, G. Shirane et al.\cite{Shirane} measured the magnetic structures of the parent compound FeCr$_2$O$_4$ by neutron powder diffraction (NPD). In 2008, Tomiyasu et al.\cite{Tomiyasuf} reported the dynamical spin-frustration effect on the magnetic excitations of FeCr$_2$O$_4$. Both composition and temperature dependences of cubic-tetragonal-orthorhombic structure transitions had been reported by X-ray powder diffraction (XPD) and specific heat.\cite{Yearian, Goodenough, Kyono, Francombe, Levinstein, Kose} The magnetic properties had been studied by magnetic and M\"{o}ssbauer effect measurements.\cite{Francombe, Robbins, Kose, Ok, Quintiliani} However, a systematic study of the doping effect on magnetic structures is still missing, and the T-x phase diagram is still under debate.\cite{Levinstein, Kose}

In this paper, the magnetic and crystal structure of Fe$_{1+x}$Cr$_{2-x}$O$_{4}$ (0.2 $\le x \le$ 1.0) are studied by NPD, XPD, magnetic susceptibility, and specific heat measurements. The phase diagram for 0.0 $\le x \le$ 1.0 is also constructed. The measurements confirm the existence of a paramagnetic-to-collinear ferrimagnetic phase transition for the entire $x$ range and a conical ferrimagnetic state at low temperature in the low Fe-doping region ($x \le$ 0.6). The structural phase transition is complicated: Although the cubic-to-tetragonal transition (0.0 $\le x \le$ 0.8) and a short-range tetragonal distortion is suggested (0.8 $\le x \le$ 1.0), the related structural transition temperature ($T_{S_1}$) decreases at $x \le$ 0.3, then increases gradually at 0.3 $\le x \le$ 0.8, and decreases again at $x \ge$ 0.8. In addition, a tetragonal-to-orthorhombic transition is detected at $T_{S_2}$ and disappears at the high Fe-doping region ($x \ge$ 0.7). We extend the study of phase diagram to compositions beyond 0.4 and temperatures below 80 K. These observations not only emphasize the competitions among the spin-lattice interaction, the Jahn-Teller distortion, the cooperative spin-orbital coupling, and the disordered states of the $A$- and $B$-site ions, but also reveal the effect of the magnetic moment magnitude and electron hopping on this frustrated spinel system.

\section{Experiment}

Polycrystalline samples of Fe$_{1+x}$Cr$_{2-x}$O$_{4}$ (0 $\le x \le$1.0) were synthesized by solid state reaction. Stoichiometric mixtures of Fe$_2$O$_3$, Fe, and Cr$_2$O$_3$ were ground together and calcined under flowing Ar at 1150$^\circ$C for 20 h. The magnetic susceptibility was measured with a SQUID (Quantum Design) with an applied field H = 100 Oe with zero field(ZFC) and field cooling processes(FC). The specific heat measurements were performed on a Quantum Design physical property measurement system (PPMS).

Low-temperature XPD patterns were collected using a PANalytical Multi-Purpose Diffractometer (MPD) equipped with an Oxford Cryosystems PheniX cryostage closed-cycle helium refrigerator. The MPD was configured with copper $K_{\alpha_{1,2}}$ radiation and fixed slits, a diffracted-beam monochromator to minimize background fluorescence from iron, and a high-speed X'celerator position-sensitive detector. The powder samples were either pressed in a stainless-steel cup or solution-cast from ethanol onto an anodized flat holder, depending on the amount of material available. The instrument alignment was verified prior to the data collection using NIST 660a LaB$_6$ standard, but no internal standards were used in order to prevent contamination of the samples. The cryostage was operated under a vacuum of approximately 10$^{-6}$ Torr. Data were collected over broad and limited diffraction angles in order to verify structure and to carefully examine the structural transitions over a broad temperature range from 15 to 300 K. The X'Pert HighScore Plus software was employed to identify possible phases and determine the lattice parameters.

NPD experiments were performed at the High Flux Isotope Reactor (HFIR) of the Oak Ridge National Laboratory (ORNL). For each composition, about 5 g of powder was loaded in a vanadium-cylinder can. A closed-cycle refrigerator was employed for samples with $x$ $\le$ 0.8, while a cryofurnace was used for samples with 0.9 $\le$ $x$ $\le$ 1.0. Preliminary neutron diffraction data were obtained from the Wide Angle Neutron Diffractometer (WAND). High-resolution neutron powder diffraction measurements were performed using the neutron powder diffractometer, HB2A. Data were collected at selected temperatures using two different wavelengths $\lambda$ =1.538 and 2.406 \AA$\, $ and collimation of 12'-open-6'. The shorter wavelength gives a greater intensity and higher Q coverage that was used to investigate the crystal structures in this low temperature regime, while the longer wavelength gives lower Q coverage with better resolution that was important for investigating the magnetic structures of the material. The diffraction data were analyzed using the Rietveld refinement program FullProf.\cite{Juan}

\begin{figure}
 \centering
  \includegraphics[width=0.45\textwidth]{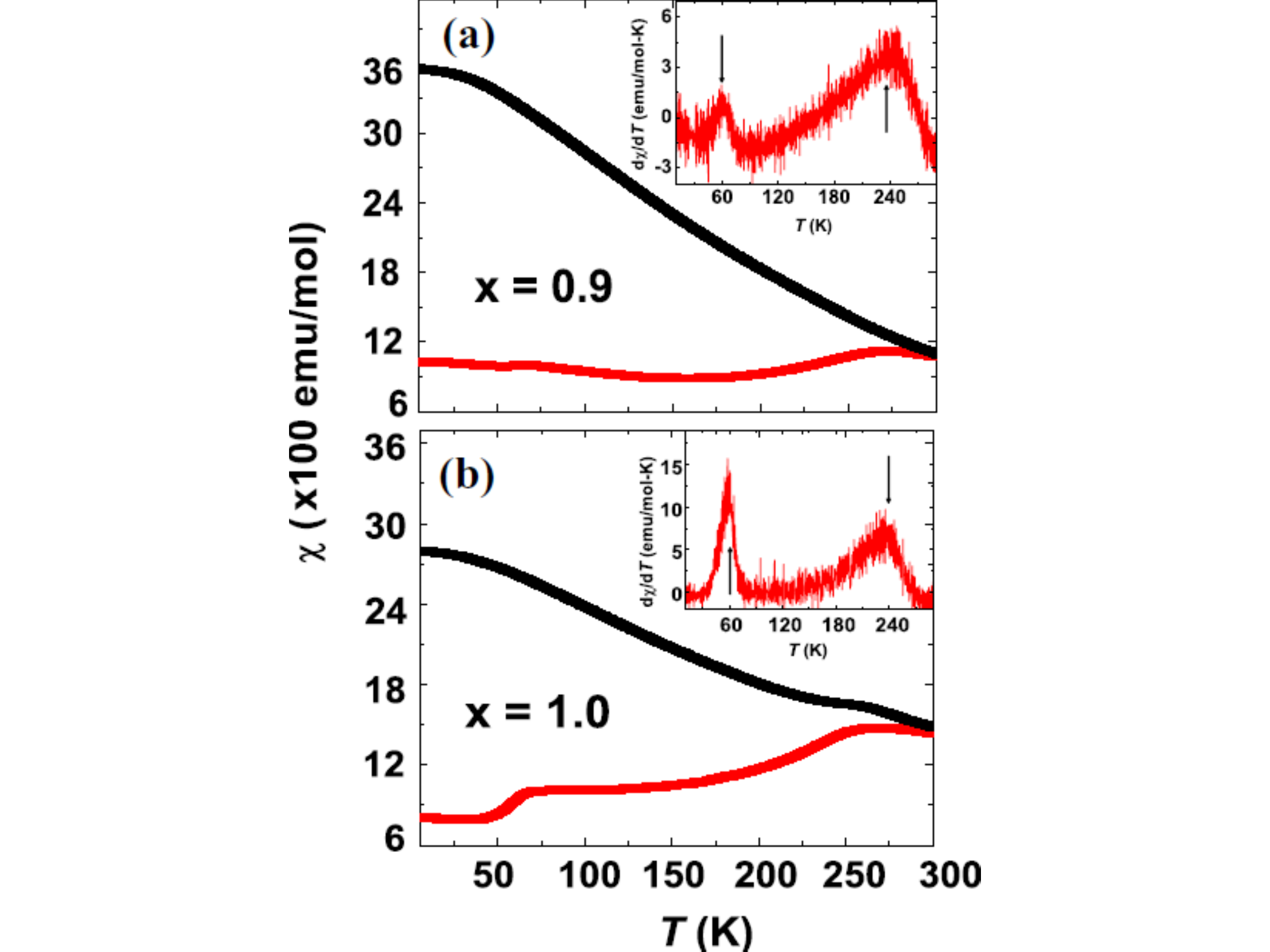}\\
\caption{(Color online) The temperature dependence of magnetic susceptibilities for Fe$_{1+x}$Cr$_{2-x}$O$_4$ ($x$ = 0.9(a) and 1.0(b)). Black and red lines are the results of FC and ZFC measurements, respectively. The insets are the related temperature derivative of the ZFC susceptibilities and the arrows mark the peak positions in temperature.}
  \label{SQUID_high}
\end{figure}

The magnetic order parameter measurements were carried out using the HB1A triple-axis spectrometer at HFIR. HB1A was operated with an incident neutron wavelength of $\lambda$ = 2.359 \AA. A pyrolytic graphite (PG) (002) monochromator and analyzer were used together with collimation of 40$^\prime$-40$^\prime$-40$^\prime$-80$^\prime$. Contamination from higher-order beams was removed using PG filters.

\section{Results}
\subsection{Magnetic Susceptibility and Specific Heat }

\begin{figure}
 \centering
  \includegraphics[width=0.575\textwidth]{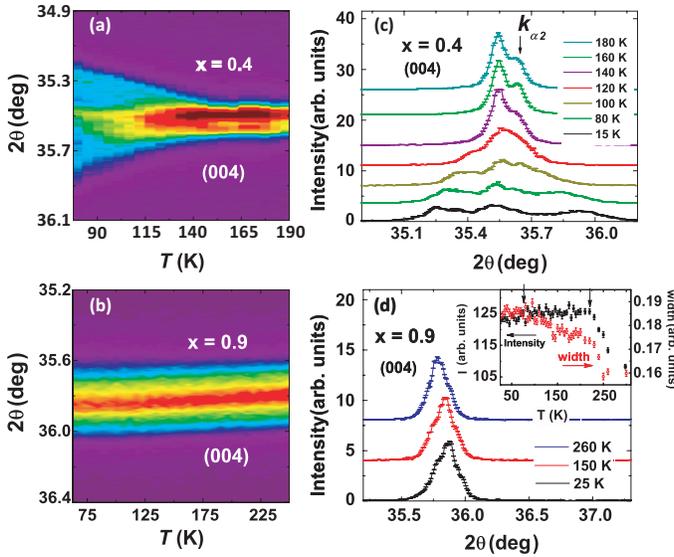}\\
\caption{(Color online) The 2$\theta$ dependence of the (004) Bragg peak of Fe$_{1+x}$Cr$_{2-x}$O$_4$ ($x$ = 0.4(a) and 0.9(b)) as a function of temperature by XPD. The XPD data around the cubic (004) Bragg position of $x$ = 0.4(c) and 0.9(d) at selected  temperatures. Inset shows the width(open circle) and integrated intensity(filled square) of the (004) Bragg reflections of $x$ = 0.9 by XPD.}
  \label{OP}
\end{figure}

Figure~\ref{SQUID} shows the temperature dependence of magnetic susceptibilities for Fe$_{1+x}$Cr$_{2-x}$O$_{4}$ (0.0$\le x \le$ 0.8). As the temperature decreases, the curves of ZFC and FC split at the paramagnetic-to-ferrimagnetic transition temperature, $T_{C}$, and an additional magnetic phase transition becomes apparent at low temperature, $T_N$, for the low Fe-doped compounds ($x \ge$ 0.6). The composition dependent $T_{C}$ agree with the previous reports.\cite{Francombe, Robbins, Kose} As Fe content increases, $T_{C}$ increases gradually. For the second magnetic phase transition temperature, $T_N$, it decreases with more Fe$^{3+}$ ions introduced to the $B$-site. As described in the following section, this transition corresponds to a spin reorientation into a noncollinear conical state.

For the high Fe-doped compounds ($x \ge$ 0.9), $T_{C}$ is above room temperature, as shown in Fig.~\ref{SQUID_high}. Although there is no obvious anomaly observed from the FC data, there are two peaks obtained from the temperature derivative of the ZFC susceptibility, which might be related to Jahn-Teller effects from the non-degenerate $e_g$ and $t_{2g}$orbitals and will be discussed in the following sections.

In order to check the effect of the magnetic transitions on the lattice, the specific heat was measured from 2 to 300 K for several compounds, such as $x$ = 0.0, 0.1, 0.4, 0.5, and 0.7, as shown in Fig.~\ref{SQUID}(a) -- (e). For $x \le$ 0.4, the structural transition temperatures are comparable to the data of Kose et al.. \cite{Kose} The apparent structural transition becomes less distinct with Fe-doping. In addition, the data of specific heat clearly present the collinear-to-noncollinear ferrimagnetic transition for $x \le$ 0.6. Therefore, there are two distinct structural transitions for the low Fe concentration ($x \le$ 0.3). One is observed in the paramagnetic state, above $T_C$, and the other emerges in the vicinity of the magnetic ordering temperature. Those structural transitions are also in agreement with the diffraction data, which we will discuss the details in the following sections.

\subsection{Neutron and X-ray diffraction}

The structural and magnetic phases of Fe$_{1+x}$Cr$_{2-x}$O$_4$ system are identified by NPD and XPD on several different compositions. The temperature dependences of the (004) Bragg peak intensity of Fe$_{1+x}$Cr$_{2-x}$O$_4$ ($x$=0.4 and 0.9) measured by XPD are plotted in Fig.~\ref{OP}.

\begin{table*} [tph]
\caption{Crystallographic information and Rietveld profile reliability factors for Fe$_{1+x}$Cr$_{2-x}$O$_4$ ($x$=0.2, 0.3, 0.6, 0.8, and 1.0) from NPD data at 5 K.}
\begin{tabular}{lccccc}
\hline
$\,  \,  \,  \,  \,  \, \,  \,  \,  \,  \,  \, \,  \,  \,  \,  \,  \, \,  \,  \,  \,  \,  \, \,  \,  \,  \,  \,  \, \,  \,  \,  \,  \,  \, \,  \,  \,  \,  \,  \, \,  \,  \,  \,  \,  \, $  & $\,  \,  \,  \,  \,  \, \,  \,  \,  \,  \,  \, $ $x$ = 0.2  $\,  \,  \,  \,  \,  \, \,  \,  \, $ & $\,  \,  \,  \,  \,  \, \,  \,  \,  \,  \,  \, $ $x$ = 0.3  $\,  \,  \,  \,  \,  \, \,  \,  \, $ & $\,  \,  \,  \,  \,  \, \,  \,  \,  \,  \,  \, $ $x$ = 0.6  $\,  \,  \,  \,  \,  \, \,  \,  \, $ & $\,  \,  \,  \,  \,  \, \,  \,  \,  \,  \,  \, $ $x$ = 0.8  $\,  \,  \,  \,  \,  \, \,  \,  \, $ & $\,  \,  \,  \,  \,  \, \,  \,  \,  \,  \,  \, $ $x$ = 1.0  $\,  \,  \,  \,  \,  \, \,  \,  \, $\\
\hline
Crystal symmetry & orthorhombic & orthorhombic & orthorhombic & tetragonal & cubic\\
Space group& $F d d d$ & $F d d d$ & $F d d d$ & $I 41/a m d$ & $F d \bar{3} m$\\
$\,  \,  \,  $\emph{a} ($\AA$)& 8.4300(4) & 8.4294(4) & 8.3945(5) & 5.9179(1) & 8.3792(1)\\
$\,  \,  \,  $\emph{b} ($\AA$)& 8.4710(3) & 8.4754(4) & 8.4919(4) & -- & -- \\
$\,  \,  \,  $\emph{c} ($\AA$)& 8.2343(3) & 8.2414(5) & 8.3172(5) & 8.4193(1) & --\\
$\,  \,  \,  $\emph{c}/\emph{a} & 0.977 & 0.978 & 0.991 & 1.001 & 1.000\\
$\,  \,  \,  $\emph{V} ($\AA$$^3$) & 588.0(1) & 588.8(1) & 592.9(1) & 294.86(1) & 588.32(1)\\
$\,  \,  \,  $\emph{Z}  &  8 &  8 &  8 &  8 &  4 \\
Recording angular range($^\circ$) & 10.5--131.9 & 10.5--131.9 & 10.5--131.9 & 10.5--131.9 & 10.5--131.9\\
calculated density ($g/cm^3$) & 5.074 & 5.068 & 5.067 & 5.110 & 5.142\\
Bragg \emph{R}-factor     &  11.5 &  8.5 &  5.7 &  5.5 &  5.6\\
Magnetic \emph{R}-factor  &  11.1 &  8.2 &  6.9 &  6.8 &  4.2\\
\hline
\end{tabular}
\label{lattice}
\end{table*}

\begin{figure}
 \centering
  \includegraphics[width=0.425\textwidth]{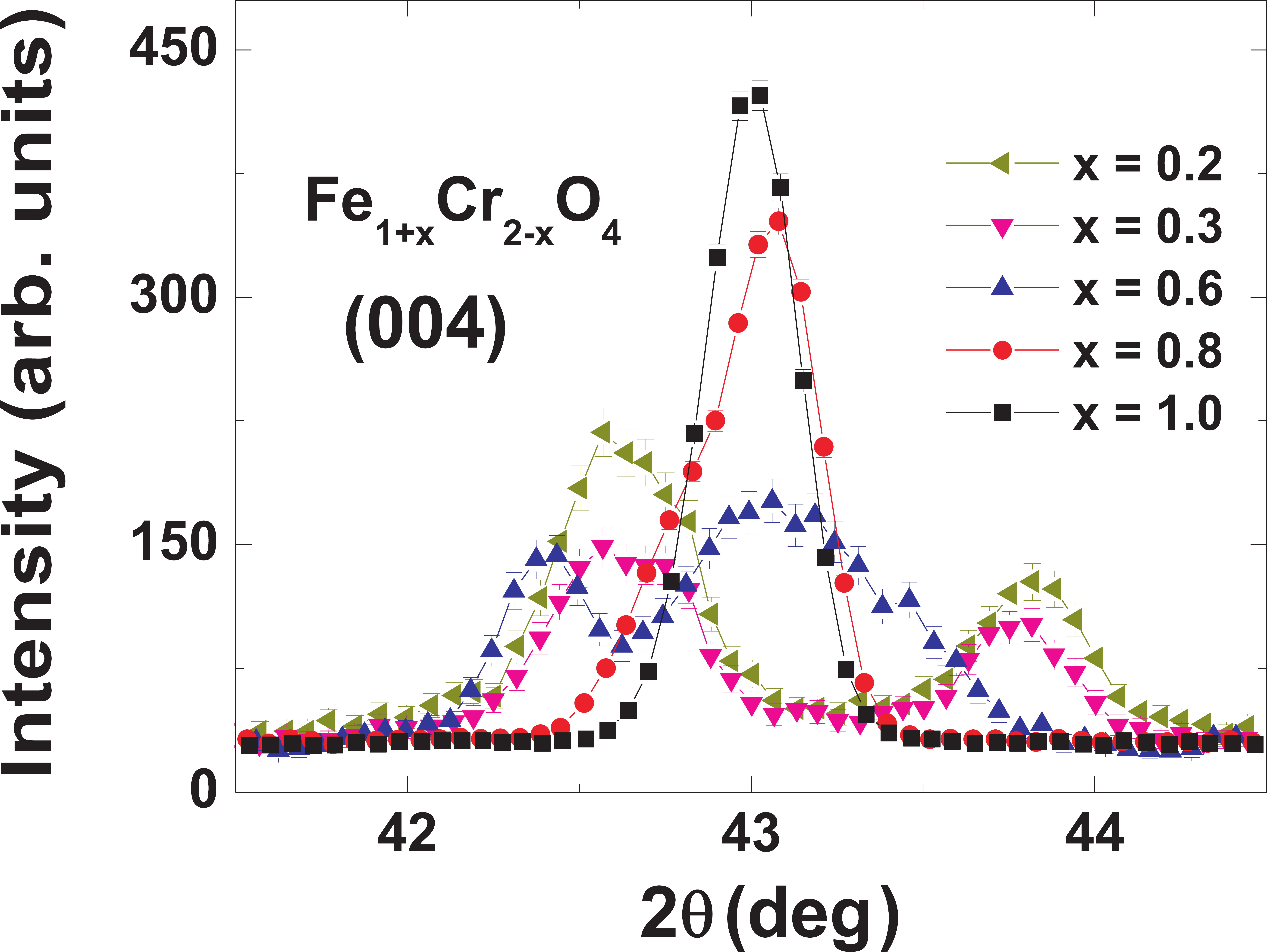}\\
\caption{(Color online) The NPD data around the cubic (004) Bragg position for different compositions at 5 K}.
  \label{NOP}
\end{figure}

For the low Fe-doped compound, such as $x = 0.4$, the structure changes from cubic to tetragonal at $T_{S_1}$ $\sim$ 150 K and then to orthorhombic at $T_{S_2}$ $\sim$ 110 K, as shown in Fig.~\ref{OP}(a); For the high Fe-doped compound, such as $x = 0.9$, there is no obvious structural phase transition observed and the symmetry is cubic, as shown in Fig.~\ref{OP}(b). The composition dependence of the diffraction pattern observed at 5 K is shown in Fig.~\ref{NOP}. The three different patterns were well described by the space groups $F d \bar{3} m$ (one peak), $I41/amd$ (two peaks), and $Fddd$ (three peaks). This is consistent with the previous analysis and reflecting the cubic-tetragonal-orthorhombic sequence of structural phase transitions upon decreasing the Fe amount.\cite{Yearian, Goodenough, Kyono, Francombe, Levinstein, Kose} Because of the limited instrumental resolution, the two peaks of the tetragonal phase ($x = 0.8$) are observed as one broad peak, and two peaks (one is sharp and the other is broad) are observed in the orthorhombic phase instead of three sharp peaks for $x = 0.2, 0.3$ and 0.6, Fig.~\ref{NOP}. For the high Fe-doped compound ($x \ge$ 0.9), although the phase transition related peak splitting is not fully perceived with the current XPD experimental resolution, the peak widths of (004) at 20 K are broader than at 300 K. The two characteristic temperatures indicated by the magnetic susceptibility measurements are marked by arrows shown in Fig.~\ref{OP}(d). The temperature dependent anomaly at $\sim$ 210K is strong on both integrated intensities and peak widths, hence a tetragonal phase is suggested, which agrees with Francombe et al..\cite{Francombe} On the other hand, there is no apparent anomaly at $\sim$ 60 K. It is possible that the anomalies in the magnetic susceptibilities originate from the transition/crossover to the different ground state of $e_g$ orbitals due to the structural transition/distortion. Therefore, we speculate that the anomaly at $\sim$ 60 K is also related with a structural distortion. Due to the instrumental resolution, the difference between a structural distortion and a structural transition cannot be fully resolved, i.e. a change that is observed as a peak broadening rather than a distinct peak splitting. So far, we cannot confirm the new phase around 60 K with lab XPD. Synchrotron or single crystal diffraction could make it possible to investigate the structural transitions/distortions more accurately, however, this is beyond the scope of the current investigations.

\begin{figure}
 \centering
  \includegraphics[width=0.48\textwidth]{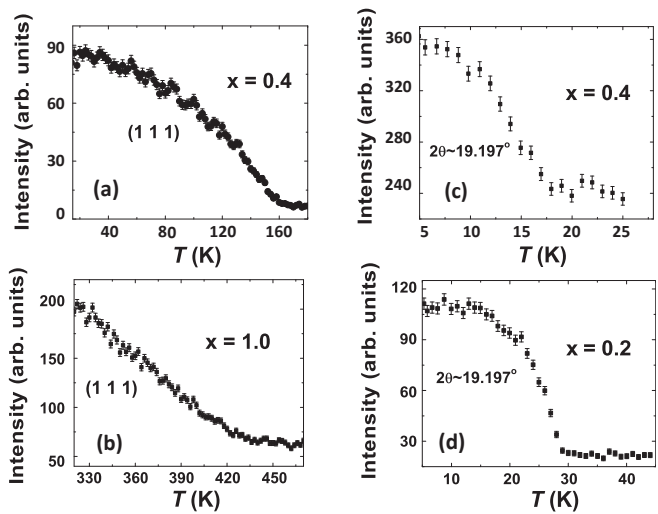}\\
\caption{(Color online) The neutron diffraction data of Fe$_{1+x}$Cr$_{2-x}$O$_4$ upon warming.  The paramagnetic-to-collinear ferrimagnetic transitions of $x$ = 0.4 (a) and 1.0 (b) are presented by the integrated intensities of (111) Bragg peaks, and the integrated intensities of the incommensurate reflection at 19.197$^\circ$ presents the collinear-to-conical ferrimagnetic transitions of $x$ = 0.4 (c) and 0.2(d), respectively.}
  \label{OP_NPD}
\end{figure}

Fig.~\ref{OP_NPD} presents the order parameters of the (111) and incommensurate satellite reflections measured by NPD for different Fe concentrations. The rise of the (111) magnetic Bragg intensities indicate a collinear ferrimagnetic order set in at $\sim$ 150 K ($x = 0.4$) and $\sim$ 410 K ($x = 1.0$), respectively, which are in a good agreement with the previous reports \cite{Francombe, Levinstein, Robbins, Kose} and the bulk magnetization well for $x = 0.4$, as shown in Fig.~\ref{SQUID}. Similar to the parent compound, FeCr$_2$O$_4$,\cite{Shirane} the collinear-to-noncollinear ferrimagnetic transitions are also observed by the appearance of incommensurate magnetic reflections at $T_N$, which are $\sim$ 28 K for $x = 0.2$ and $\sim$ 18 K for $x = 0.4$, respectively. Actually, this incommensurate peaks have also been reported in other magnetic $A$-site chromites, such as CuCr$_2$O$_4$, MnCr$_2$O$_4$ and CoCr$_2$O$_4$.\cite{Hastings, Menyuk, Plumier, Tomiyasu, Ederer, Ohtani}

\begin{figure}
 \centering
  \includegraphics[width=0.48\textwidth]{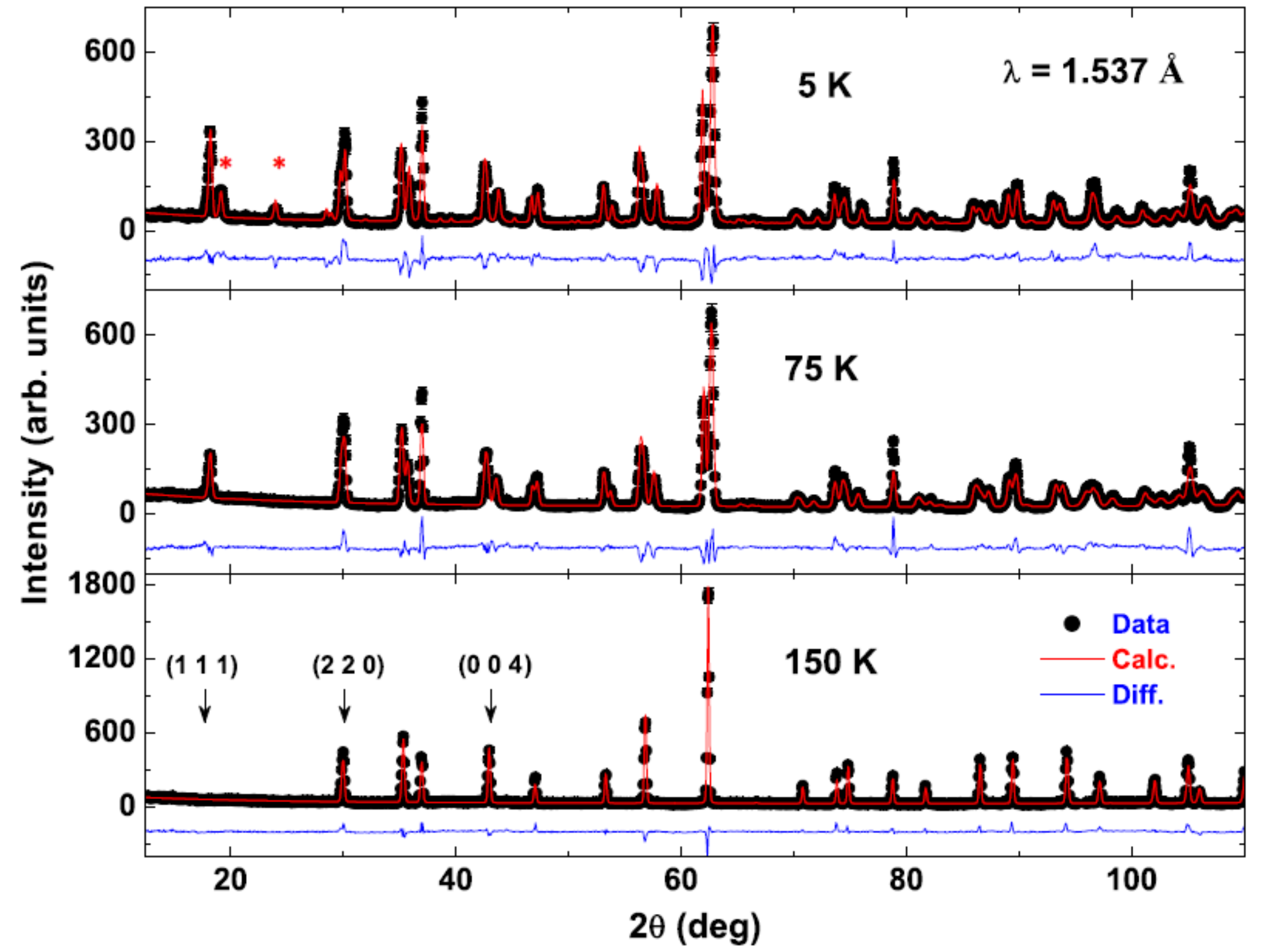}\\
\caption{(Color online) Plots of raw NPD data (black dots) for Fe$_{1.2}$Cr$_{1.8}$O$_4$ measured at $T$ = 5, 75, and 150 K for 2$\theta$ $\le$ 120$^{\circ}$. Solid lines are results of Rietveld refinements described in the main text. Differences between observed and calculated intensities are shown directly below the respective patterns. The stars indicate an incommensurate phase.}
  \label{diff_HB2A}
\end{figure}

\begin{figure}
 \centering
  \includegraphics[width=0.45\textwidth]{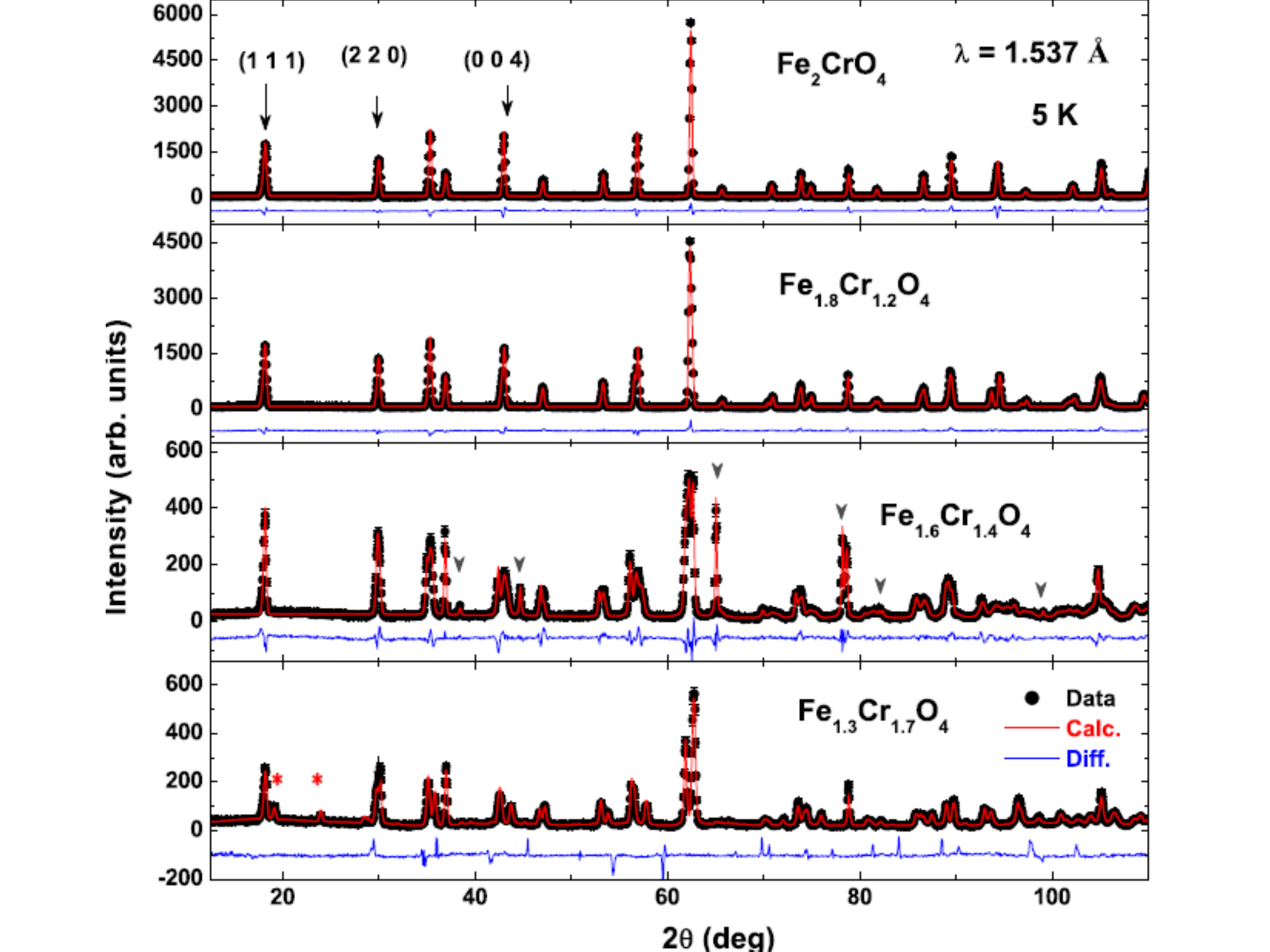}\\
\caption{(Color online) Plots of NPD data (black dots) for Fe$_{1+x}$Cr$_{2-x}$O$_4$($x$ = 0.3, 0.6, 0.8, and 1.0) measured at $T$ = 5 K for 2$\theta$ $\le$ 120$^{\circ}$. Solid lines are results of Rietveld refinements described in the main text. Differences between observed and calculated intensities are shown below the respective patterns. The stars indicate an incommensurate phase and the grey arrows are the signal from Al-can.}
  \label{HB2A_5K}
\end{figure}

Rietveld analyses were employed to determine precisely the changes in both the crystal and magnetic structures for each composition. The Rietveld fitted patterns on Fe$_{1.2}$Cr$_{1.8}$O$_4$ at 150, 75, and 5 K are shown in Fig.~\ref{diff_HB2A}. Figure~\ref{HB2A_5K} displays the Rietveld profile fitting results at 5 K for selected Fe compositions.  The data at 150 K for Fe$_{1.2}$Cr$_{1.8}$O$_4$ confirm the cubic spinel structure(defined by $F d \bar{3} m$ space group) without impurity phases. The refinement results indicate that less than 0.5$\%$ dislocations between \emph{A}- and \emph{B}-site in Fe$_{1.2}$Cr$_{1.8}$O$_4$, which confirms the statement of the Cr$^{2+}$ energy level lying well above the Fe$^{2+/3+}$ energy level.\cite{Yearian, Goodenough} Similar to FeV$_2$O$_4$,\cite{MacDougall} the diffraction patterns are well described by the space groups $Fd\bar{3}m$ for the cubic lattice and the $Fddd$ for orthorhombic lattice with decreasing temperature, while the magnetic phases are well described by the collinear and conical magnetic states, respectively. The conical magnetic phase is an incommensurate phase (with the corresponding reflections labeled by stars in Fig.~\ref{diff_HB2A}(a) and Fig.~\ref{HB2A_5K}(a)). Upon Fe-doping, the positions of incommensurate peaks do not change significantly, but the intensities decrease. In order to model these reflections, we have tried the conical model with ferrimagnetic order along [110] (as in MnCr$_2$O$_4$\cite{Hastings, Tomiyasu}) or along \emph{c}-axis (as found for CuCr$_2$O$_4$\cite{Prince}). The quality of the fits is not very satisfactory for either model, being affected by an anisotropic peak broadening which may be results from the peak broadening which might come from the microstrains or other structural distortions in the sample. One the other hand, the lack of enough unique magnetic peaks in this powder data hinders the reliable determination of the direction and magnitude of the Fe$^{2+}$ and Fe$^{3+}$/Cr$^{3+}$ magnetic moments. Single crystal neutron diffraction measurements are clearly needed to determine the exact canting angles and ordered moments. Figs.~\ref{diff_HB2A}(a) and ~\ref{HB2A_5K}(a) present the best fits from the refinements with the propagation vector of $k$ = [0.391, 0.391,0] for the centered cell $Fddd$. Detailed information about the structural refinement and the atomic coordinates is summarized in Table~\ref{lattice} and ~\ref{positions}. The doping effect on the tetragonal splitting is also confirmed by the change of the $c/a$ ratio.\cite{Shirane, Yearian, Goodenough, Kyono, Francombe, Levinstein, Kose}


The cubic-tetragonal-orthorhombic sequence of structural phase transition is also captured by the XPD measurements with 5 K/ step, which is consistent with the previous analysis\cite{Kose} and similar to other Fe$B_2$O4 spinels, such as Mn-doped chromite Fe$_{1-x}$Mn$_{x}$Cr$_{2}$O$_4$ \cite{Ohtani} and Fe-vanadate Fe$_{1+x}$V$_{2-x}$O$_4$.\cite{MacDougall, Lee, Liu}

$\\$
\section{Discussion}

\begin{figure}
 \centering
  \includegraphics[width=0.48\textwidth]{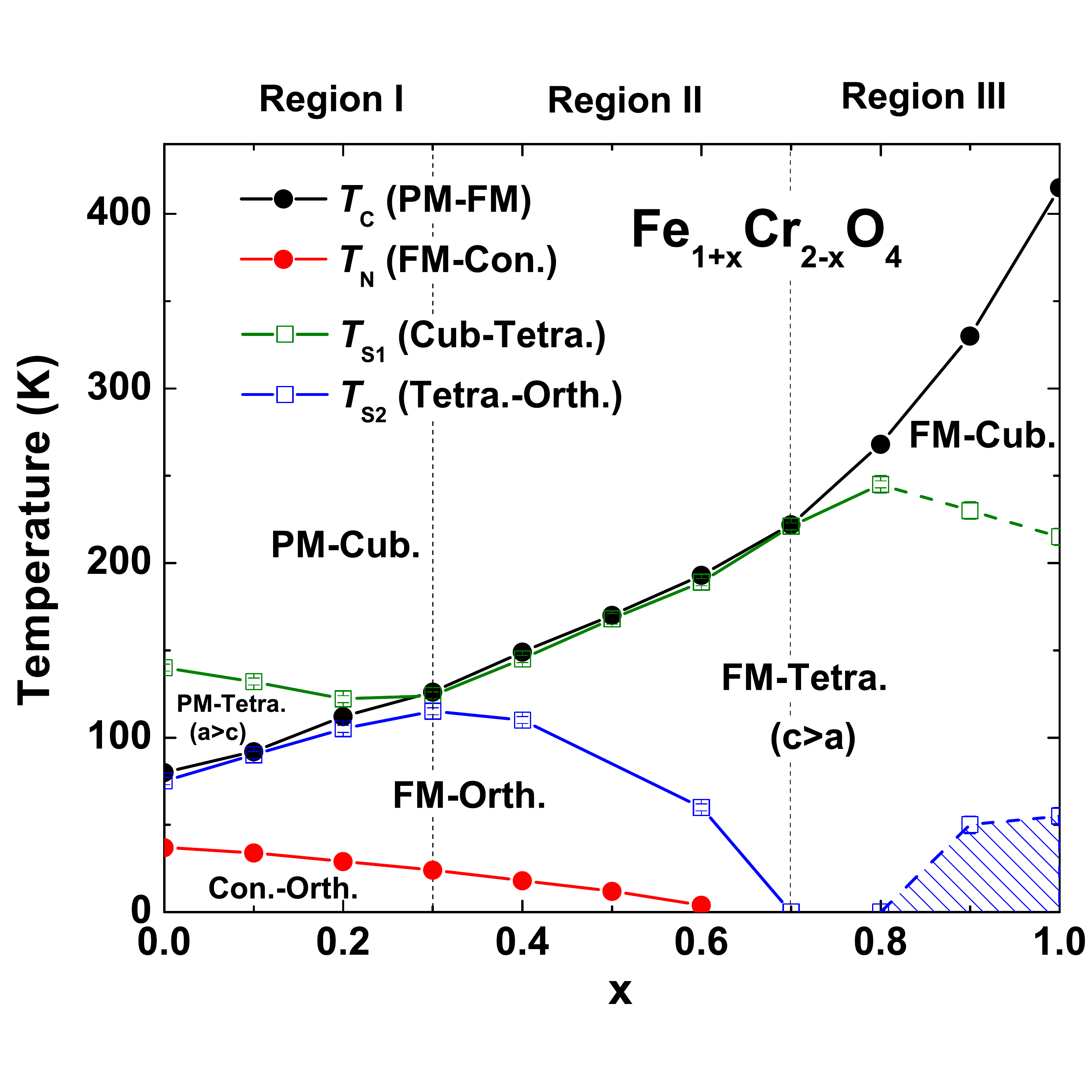}\\
\caption{(Color online) The temperature versus Fe content ($x$) phase diagram of Fe$_{1+x}$Cr$_{2-x}$O$_{4}$. $T_C$ is the paramagnetic-to-collinear ferrimagnetic phase transition temperature(black lines and dots), $T_{N}$ is the collinear-to-conical ferrimagnetic phase transition temperature(red lines and dots), $T_{S_1}$ is the cubic-to-tetragonal lattice transition temperature(olive line with open squares), and $T_{S_2}$ is the tetragonal-to-orthorhombic lattice transition temperature(blue line with open squares). The solid lines display the structural and magnetic transitions, while the dashed lines display the possible structural distortions.}
  \label{phase}
\end{figure}

Combining the diffraction and the magnetic susceptibility measurements, a $T-x$ phase diagram including both the crystal and magnetic structures can be constructed as shown in Fig.~\ref{phase}. The complicated phase diagram clearly presents the Fe$^{3+}$-doping effects on the Jahn-Teller distortion, spin-lattice interaction, orbital-lattice interaction, Fe$^{2+}$-Fe$^{3+}$ hopping, and disordering effect of Fe$^{2+}$, Fe$^{3+}$, and Cr$^{3+}$ ions in the system. There are three major regions, which will now be discussed separately.

$\\$
i) $x$ $\le$ 0.3,
$\\$

The doped Fe$^{3+}$ ions only occupy the $B$-site of the spinel, and the compounds have the normal type of structure with the formula Fe$^{2+}$[Cr$^{3+}_{2-x}$Fe$^{3+}_x$]O$_4$.\cite{Yearian, Goodenough, Kyono, Francombe, Levinstein, Robbins, Kose, Quintiliani} The paramagnetic-to-collinear ferrimagnetic and collinear-to-conical ferrimagnetic phase transitions are observed at $T_{C}$ and $T_{N}$, respectively. With increasing Fe-doping, $T_{C}$ increases and $T_{N}$ decreases. Although the cubic, tetragonal, and orthorhombic phases are observed in sequence as temperature decreases, $T_{S_1}$ decreases with Fe-doping while $T_{S_2}$ increases with $T_{C}$. Moreover, the lattice constant $a$ is larger than $c$ in the related tetragonal phase.

The 6 outer-shell electrons of Fe$^{2+}$ occupy the 3$d$ orbitals ($e_g^3t_{2g}^3$), giving one of the three $e_g$ electrons the orbital degree of freedom on 3$d_{z^2}$ or 3$d_{x^2-y^2}$. Since the FeO$_4$ tetrahedra are generated from a cube where one Fe$^{2+}$ ion is located at the center of four O$^{2-}$ ions that occupy two diagonal corners, the distortion modes can be represented, as discussed in Ref. [34], by a combination of 3$d_{z^2}$ and 3$d_{x^2-y^2}$, which are described by $Q_2$ and $Q_3$, respectively,\cite{Vleck, Ohtani, Opik}

\begin{equation}
   \begin{split}
Q_2 &= \frac{1}{\sqrt{2}L} (\delta X - \delta Y) \, , \\
Q_3 &= \frac{1}{\sqrt{6}L} (2\delta Z - \delta X- \delta Y) \, ,
   \end{split}
\label{Q23}
\end{equation}
where $L$ is the length of the related cube, $\delta X$, $\delta Y$, and $\delta Z$ are the modulation of the cube dimensions.

Thus, the Hamiltonian of the coupling between the distortion and orbital occupation should be,

\begin{equation}
    H = -A (\tau_x Q_2 + \tau_z Q_3) \, ,
\label{Hamiltonian}
\end{equation}
where $\tau$ is the Pauli matrix and $A$ is the coupling constant.

As discussed by \"{O}pik and Pryce,\cite{Opik} the total of the orbital-lattice coupling in Eq.(~\ref{Hamiltonian} ) and the $Q_2$ ($Q_3$) related harmonic potential energy can be minimized by an infinite number of distortions, however, the orbital degeneracy can be lifted by the anharmonic lattice potential term of the total potential energy,
\begin{equation}
    V = \frac{1}{2}M \omega^2 Q^2 +A_3 Q^3 cos3\theta + \cdots \, ,
\label{V}
\end{equation}
where $Q$ and $\theta$ are the polar coordinations for $Q_2-Q_3$ space. $A_3$ is the term describing the anharmonic effect on the tetragonal distortion. If $A_3 >$ 0, the complex is compressed along the tetragonal axis; If $A_3 <$ 0, the complex is elongated along the tetragonal axis.\cite{Opik, Ohtani}

For the low Fe-doping FeO$_4$ tetrahedron ($x \le$ 0.3), the $e_g$ orbital shape is deduced to be of 3$d_{z^2}$ type in the paramagnetic phase, thus $A_3$ is positive and $a/c >$ 1, as shown in Fig.~\ref{phase}. As more empty $e_g$ Cr$^{3+}$ ions are replaced by the half filled $e_g$ Fe$^{3+}$ ions, the Jahn-Teller distortion of Fe$^{2+}$ ions becomes unstable and $T_{S_1}$ decreases gradually.

As temperature decreases, the ferrimagnetic state is reached and the effect of spin-orbit coupling needs to be included in the total Hamiltonian. Although the first-order perturbation of spin-orbit coupling is absent, the second-order term $\lambda {\bf L} \cdot {\bf S}$ breaks the degeneracy of the two $e_g$ orbitals and lowers the energy of the 3$d_{x^2-y^2}$ relative to the 3$d_{z^2}$ orbital.\cite{Opik, Ohtani, Kanamori} The second-order perturbation of the Hamiltonian $H_{SO}$ can be presented as,

\begin{equation}
    H_{SO} = \frac{B}{6}((3S_z^2-S^2)\tau_z-\sqrt{3}(S_x^2-S_y^2)\tau_x) \, ,
\label{V}
\end{equation}
where $B$ is the energy difference between 3$d_{x^2-y^2}$ and 3$d_{z^2}$ states.

\begin{table} [tph]
\caption{Refined atomic positions of Fe$_{1+x}$Cr$_{2-x}$O$_4$ ($x$=0.2, 0.3, 0.6, 0.8, and 1.0) from NPD data at 5 K.}
\begin{tabular}{lccccc}
\hline
$\,  \,  \,  \,  \,  \, \,  \,  \,  \,  \, \,  \,  $  & $\,  \,  \, \,  \,  \,  $ $atoms$  $\,  \,  \,  $ & $\,  \,  \,  \, $ $site$  $\,  \,  \,  \,  $ & $\,  \,  \, \,  \,  \,  \,  \,  \, $ $x$  $\,  \,  \,  \,  \,  \, $ & $\,  \,  \, \,  \,  \,  \,  \,  \, $ $y$  $\,  \,  \,  \,  \,  \, $& $\,  \,  \, \,  \,  \,  \,  \,  \, $ $z$  $\,  \,  \,  \,  \,  \, $ \\
\hline
$x$ =0.2 &Fe(1)& 8a &0.125 &0.125 &0.125\\
&Fe(2) & 16d &0.5 & 0.5&0.5\\
&Cr & 16d & 0.5 & 0.5 & 0.5\\
&O & 32e & 0.261(2) & 0.265(2) & 0.259(2) \\
\hline
$x$ =0.3 &Fe(1) & 8a &0.125 &0.125 &0.125\\
&Fe(2) & 16d &0.5 & 0.5&0.5\\
&Cr & 16d & 0.5 & 0.5 & 0.5\\
&O & 32e & 0.262(2) & 0.265(2) & 0.258(2) \\
\hline
$x$ =0.6 &Fe(1) & 8a &0.125 &0.125 &0.125\\
&Fe(2) & 16d &0.5 & 0.5&0.5\\
&Cr & 16d & 0.5 & 0.5 & 0.5\\
&O & 32e & 0.263(2) & 0.265(2) & 0.260(2) \\
\hline
$x$ =0.8 &Fe(1)&4a &0.0 &0.75 &0.125\\
&Fe(2) & 8d &0.0 & 0.0&0.5\\
&Cr & 8d & 0.0 & 0.0& 0.5\\
&O & 16h & 0.0& 0.020(2)& 0.262(1) \\
\hline
$x$ =1.0 &Fe(1)& 8a &0.125 &0.125 &0.125\\
&Fe(2) & 16d &0.5 & 0.5&0.5\\
&Cr & 16d & 0.5 & 0.5 & 0.5\\
&O & 32e & 0.259(2) & 0.259(2) & 0.259(2) \\
\hline
\end{tabular}
\label{positions}
\end{table}

Thus, the sign of $A_3$ is changed to negative to form the orthorhombic phase at low temperature, and $T_C$ is above the associated structural transition temperature $T_{S_2}$.\cite{Opik, Ohtani}

At the same time, the magnetic transition temperature could be roughly estimated by mean-field-theory,
\begin{equation}
    3 k_{B} T_C = z\sum_{i,j}J_{ij} {\bf S_i} \cdot {\bf S_j} \, ,
\label{V}
\end{equation}
where $z$ is the number of nearest neighbors, ${\bf S}$ and $J_{ij}$ are the related moment and exchange energy, respectively.

Since the extra half-filled $e_g$ electrons of the doped Fe$^{3+}$ ions increase not only the interaction between $A$- and $B$-site ions ($J_{AB}$), but also the total moment of $B$-site ions($S_{Fe^{3+}/Cr^{3+}}$), $T_C$ increases with the doping-amount of Fe$^{3+}$ ions, which drives $T_{S_2}$ to increase. Compared to the decreasing $T_{S_1}$, they meet at around $x$ = 0.3. Hence, $x$ = 0.3 is also the boundary of the two tetragonal phases with different $c/a$, Fig.~\ref{phase}.

As in the other spinel compounds with magnetic $A^{2+}$ ions, MnCr$_2$O$_4$ and CoCr$_2$O$_4$, a conical magnetic state is also observed in the low Fe-doped FeCr$_2$O$_4$ at the lower temperature due to the geometrical magnetic frustration.\cite{Hastings, Menyuk, Plumier, Tomiyasu, Ederer} Furthermore, the transition temperature $T_N$ decreases with Fe$^{3+}$-doping. Lyons et al. \cite{Lyons} had presented that the conical state is complicated and deduced the structure from a factor of u, which is closely related to the properties of both moments and interactions between the $A-$ and $B-$site cations, 4$J_{BB}S_B$/3$J_{AB}S_A$. They presented that the conical state is stable as 8/9 $\le u \le$1.298. This factor is possibly related with the decreases of $T_N$. However, it is hard to obtain the $u$-value for Fe$_{1+x}$Cr$_{2-x}$O$_4$ quantitatively because of lack of information on the exchange energies, although $S_B$, $J_{BB}$, and $J_{AB}$ are increasing with Fe-doping. Inelastic neutron scattering measurements using single crystal are needed to clarify the statement as MnV$_2$O$_4$.\cite{Ovi} Another possible reason would be that the extra Fe$^{3+}$ on the $B-$site disturbs the homogeneous frustrated interactions and decreases $T_N$.

$\\$
ii) 0.3 $<$ $x$ $\le$ 0.7,
$\\$

As in Region I ($x$ $\le$ 0.3), the doped-Fe$^{3+}$ ions occupy the $B$-site of the spinel with the normal spinel structure in Region II, Fe$^{2+}$[Cr$^{3+}_{2-x}$Fe$^{3+}_x$]O$_4$. Although the tetragonal-to-orthorhombic phase transition still follows the cubic-to-tetragonal transition, the lattice constant ratio, $a/c$, of the tetragonal phase is less than 1. In addition, the driving forces of the two structural transitions are the reverse of those in Region I, which means that the spin-orbital coupling effect on the Jahn-Teller distortion leads to the cubic-to-tetragonal transition, while the tetragonal-to-orthorhombic transition is due to the $B$-site disorders of Fe$^{3+}$ and Cr$^{3+}$ ions. Thus, $T_{S_1}$ increases with Fe-doping, and $T_{S_2}$ decreases to 0 as $x$ approaches 0.7. For the magnetic ordering transitions, $T_{C}$ accompanies the first structural distortion ($T_{S_1}$), which increases linearly with Fe-doping. The spin reorientation transition, occurring at $T_{N}$, continues to decrease and disappears at $x$ = 0.6.

$\\$
iii) 0.7 $<$ $x$ $\le$ 1.0,
$\\$

Unlike Region I and II, the doped-Fe$^{3+}$ ions begin to occupy the $A$-site, and the Fe$^{2+}$ ions move to the $B$-site of the spinel at $x \ge$ 0.7, which makes the system very complicated, Fe$^{2+}_{1.7-x}$Fe$^{3+}_{x-0.7}$[Cr$^{3+}_{2-x}$Fe$^{3+}_{0.7}$Fe$^{2+}_{x-0.7}$]O$_4$. As presented in Fig.~\ref{phase}, $T_{C}$ increases more sharply than the linear relationship in region II due to the electron hoping effect between $A$- and $B$-site Fe$^{2+}$/Fe$^{3+}$ ions, which was confirmed by the reported M\"{o}ssbauer measurement.\cite{Robbins, Quintiliani}

The $T_{S_1}$ continues the ascending trend for $x$ up to 0.8, after which it disappears. Nevertheless, some structural distortions seem to persist up to the high-doping region (0.8 $\le x \le$1) , as evidenced by the anomalies of the magnetic susceptibility measurements as shown in Fig.~\ref{SQUID_high}, as well as by the different width of the (004) Bragg peak between 20 and 300 K, as shown in Fig.~\ref{OP}(d). However, the magnetic ordering and structural distortion temperature are disconnected  and the $T_C$ rised steeply. If we still use $T_{S_1}$ to label the temperature of the tetragonal distortion, it decreases due to the electron hopping effect on the orbitals of the Fe$^{2+}$ and Fe$^{3+}$ ions, as more Fe$^{3+}$ ions are introduced in the system, indicated by the green dashed line in Fig.~\ref{phase}. Moreover, another structural distortion is observed at lower temperature, and it is suggested to be the orthorhombic/monoclinic distortion related to the extra Jahn-Teller active Fe$^{2+}$ on the $B$-site, which is related to the $t_{2g}$ orbital freedom found in inverse spinel Fe$_3$O$_4$.\cite{Wright} $T_{S_2}$ still describes this distortion and is presented as the shaded region in Fig.~\ref{phase}.

\section{Conclusion}

The structural and magnetic  phase diagram of Fe$_{1+x}$Cr$_{2-x}$O$_4$ is investigated by means of magnetization, specific heat, x-ray and neutron scattering measurements. The substitution of Fe$^{3+}$ for Cr$^{3+}$ enhances the paramagnetic-to-collinear ferrimagnetic transition temperature $T_C$ and reduces the collinear-to-conical ferrimagnetic transition temperature $T_N$, which is likely due to the complicated interactions between $A$- and $B$-site ions.

Systematic changes in the crystal structure with temperature and composition are observed. In the low Fe$^{3+}$-doped compound ($x \le$ 0.7), both cubic-to-tetragonal and tetragonal-to-orthorhombic transitions are driven by the Jahn-Teller distortion and the related spin-orbital couplings. At $x \le$ 0.3, the magnetic energy stabilizes the orthorhombic phase to increase $T_{S_2}$, while the disorder of the $B$-site ions(Fe$^{3+}$/Cr$^{3+}$) leads to the decreasing $T_{S_1}$; at 0.3 $< x \le$ 0.7, the magnetic energy increases $T_{S_1}$, while the disorder of the $B$-site ions decreases $T_{S_2}$. In the high Fe$^{3+}$-doped compound ($x >$ 0.7), a strong electron hoping mechanism between Fe$^{2+}$ and Fe$^{3+}$ ions lead to the orbital-active Fe$^{2+}$ ions occupying both $A$- and $B$-site of the spinel, and the related $e_{g}$ and $t_{2g}$ orbital effects are observed, which results in the temperature dependence of the lattice distortions.

\begin{center}
$\textbf{ACKNOWLEDGEMENT}$
\end{center}
The research at the High Flux Isotope Reactor, Oak Ridge National Laboratory was sponsored by the Scientific User Facilities Division (J.M., M.M., C.D.D.C., O.G., S.C., W.T., A.A.A., S.X.C.) and Center for Nanophase Materials Sciences (A.R.), Office of Basic Energy Sciences, US Department of Energy. R.S. and H.D.Z. thanks for the support from JDRD program of University of Tennessee. Work at FSU is supported in part by NSF-DMR 1005293, and carried out at the National High Magnetic Field Laboratory, supported by the NSF, the DOE, and the State of Florida.

\bibliographystyle{apsrev}

\end{document}